\documentclass[aps, prd, preprintnumbers, superscriptaddress, nofootinbib, amssymb, notitlepage, eqsecnum, 10pt]{revtex4-2}

\usepackage{here}
\usepackage[dvipdfmx]{graphicx}
\usepackage{amsmath,amsthm,amssymb}
\usepackage{bm}
\usepackage{color}
\usepackage{comment}

\definecolor{cadmiumgreen}{rgb}{0.0, 0.42, 0.24}

\usepackage[scr=boondox]   
           {mathalpha}

\usepackage{amsfonts}
\usepackage{dcolumn}
\usepackage{hyperref}
\allowdisplaybreaks[1]   
\usepackage{stackengine}

\newcommand{\Mpl}{M_{\rm Pl}}

\begin{document}

\preprint{YITP-25-16}

\title{$f(T)$ Gravity: Background Dependence and Propagating Degrees of Freedom}

\author{Valentina Danieli}

\affiliation{SISSA International School for Advanced Studies, Via Bonomea 265, 34136, Trieste, Italy}
\affiliation{INFN - Sezione di Trieste, via Valerio 2, 34127, Trieste, Italy}
\affiliation{IFPU, Institute for Fundamental Physics of the Universe, Via Beirut 2, 34014 Trieste, Italy}
\affiliation{CEICO, FZU–Institute of Physics of the Czech Academy of Sciences, Na Slovance 2, 18200 Praha 8, Czech Republic}

\author{Antonio De Felice}

\affiliation{Center for Gravitational Physics and Quantum Information, 
Yukawa Institute for Theoretical Physics, Kyoto University, 
606-8502, Kyoto, Japan}

\date{\today}

\begin{abstract}
The standard cosmological model, rooted in General Relativity (GR), has achieved remarkable success, yet it still faces unresolved issues like the nature of dark matter, dark energy, and the Hubble tension. These challenges might imply the need for alternative gravitational theories. Teleparallel gravity offers a compelling framework by reformulating the gravitational interaction using torsion, rather than curvature, as its fundamental geometrical property. This paper delves into $f(T)$ gravity, an extension of the Teleparallel Equivalent of General Relativity (TEGR), which introduces non-linear modifications of the torsion scalar $T$. We focus on the role of spacetime-dependent Lorentz transformations in the vierbein formalism, examining their impact on both background solutions and perturbation dynamics. Special attention is given to the homogeneous and isotropic FLRW spacetime, as well as the anisotropic Bianchi I spacetime. Furthermore, the analysis of the propagating degrees of freedom on these spacetimes is performed. While it is well established that TEGR reproduces the same results as GR, the propagating degrees of freedom in its non-linear extension, $f(T)$ gravity, is still debated in the literature. In this work, we find that only two fields propagate in the gravity sector, independently of the background spacetime considered, either FLRW or Bianchi I. Although not definitive, this paper provides fresh insights into the issue of the propagating degrees of freedom in $f(T)$ gravity, opening the door to intriguing new directions for further investigation.
\end{abstract}

\maketitle

\section{Introduction}
\label{sec:intro}

The standard cosmological model is based on General Relativity (GR), the theory of gravity that best agrees with observations. It explains the theory of classical black holes, it predicts the production and propagation of gravitational waves, and it has been tested in a large number of situations that range from laboratory up to astrophysical distances \cite{Asmodelle:2017sxn}.
However, some mysteries persist at both very large and very small scales. Already at the Galactic scales, discrepancies arise between the motion of stars and the gravitational field generated by visible matter. This inconsistency is most likely due to the existence of dark matter \cite{Bertone:2004pz, Arbey:2021gdg}; nevertheless, Newton's law has not been directly tested at those scales and there is then the possibility that this effect might partially stem from a modification of the laws of gravity.
At cosmological scales the mismatch between the expansion due to the known matter and the observed expansion is even more dramatic, requiring that a large fraction of the energy-momentum comes from a dark energy \cite{Li:2012dt}. Furthermore, a deviation is observed between the value of the Hubble parameter $H_0$ obtained from the Planck satellite data as predicted by the $\Lambda$CDM model and the value estimated using the standard candles in the SH0ES experiment \cite{Hu:2023jqc}. If the systematics of these experiments are under control, General Relativity would be ruled out. These are some of the reasons why investigating alternative gravity theories is worthwhile.

Originally, the action for general relativity was built up using only the Ricci scalar $R$ defined as the contraction of the Riemann tensor, whose definition is given in terms of the metric $g_{\mu\nu}$. However, this choice was not the only one possible, it was just the simplest \cite{CANTATA:2021asi}. The basic principles of relativity require simply that the space-time structure has to be determined by either one or both of two fields, a metric $g$ and a linear connection $\Gamma$. With these two geometrical quantities two other mathematical objects can be constructed, besides the Riemann tensor: the torsion tensor $T^\alpha{}_{\mu\nu}$ (the antisymmetric part of the linear connection) and the non-metricity tensor $Q_{\alpha\beta\gamma}$ (the covariant derivative of the metric). Both $T^\alpha{}_{\mu\nu}$ and $Q_{\alpha\beta\gamma}$ vanish in General Relativity since the connection is assumed to be symmetric and compatible with the metric. With these new mathematical objects, many alternative theories of gravity can be built up. While the Riemann tensor gives the rotation of a vector transported on a closed curve, the torsion represents the non-closure of parallelograms formed when two vectors are transported along each other and the non-metricity gives the variation of a vector's length when transported. The gravitational interaction can be equivalently described by these three different properties of spacetime \cite{BeltranJimenez:2019esp, Heisenberg:2018vsk, Capozziello:2022zzh}. In particular, we can build an action with the Ricci scalar $R$ for GR, the torsion scalar $T$ for Teleparallel Equivalent of General Relativity (TEGR) \cite{Aldrovandi:2013wha, Maluf:2013gaa}, or the non-metricity scalar $Q$ for Symmetric Teleparallel Equivalent of General Relativity (STEGR) \cite{Nester:1998mp}. It can be proved that both $Q$ and $T$ are equivalent to the Ricci scalar $R$ up to a boundary term \cite{BeltranJimenez:2019esp}. Therefore, a theory where the action is set by the torsion scalar or the non-metricity scalar gives equivalent field equations to GR. However, this equivalence is broken when non-trivial functions of $T$ or $Q$ are considered, analogously to the $f(R)$ extensions of General Relativity \cite{Sotiriou:2008rp, DeFelice:2010aj}.
This paper aims to deepen our understanding of a specific set of theories, called Teleparallel Gravity \cite{Hohmann:2022mlc}, which are built setting to zero the Riemann tensor and the non-metricity while keeping only the torsion tensor as the fundamental building block of the theory. Significant efforts have been invested in exploring the theoretical properties and observational consequences of these theories \cite{Wu:2010mn, Fu:2011zze, Bahamonde:2021gfp, Cardone:2012xq, Farrugia:2016xcw, Chen:2019ftv, Golovnev:2020las, Ren:2021uqb, Ren:2021tfi, Golovnev:2021htv, DeBenedictis:2022sja, Zhao:2022gxl, BeltranJimenez:2021kpj, Duchaniya:2022rqu, Aljaf:2022fbk, Huang:2022slc,dosSantos:2021owt, Myrzakulov:2012axz, Wang:2011xf, Deliduman:2011ga, Arcos:2004tzt, Souza:2024qwd, Akarsu:2024nas}.

In Teleparallel Gravity the fundamental object of the theory is the vierbein $e_a=e_a{}^\mu \partial_\mu$, an orthonormal basis in the tangent space at each point of the spacetime, with $a$ an index in the tangent bundle. In this framework, the metric tensor is expressed in terms of the vierbein itself. Together with the vierbein, there is another object describing the theory: a general spacetime-dependent Lorentz transformation (LT). Indeed, $e^{a}{}_{\mu}$ is not unique as it can be boosted and rotated by a Lorentz transformation.
Notice that by definition, the metric is invariant under a LT; therefore, the Einstein-Hilbert action of General Relativity is invariant too. However, this is not the case for $f(T)$ gravity\footnote{Here $T$ is a four-dimensional scalar that corresponds to a particular linear combination of traces of the torsion tensor $T^\alpha{}_{\beta\gamma}$.}  \cite{Ferraro:2006jd, Ferraro:2008ey}, where the Lorentz transformation plays a non-trivial role in the phenomenology of the theory (cosmological background dynamics with non-trivial phenomenology can be obtained, see e.g.\ \cite{Awad:2017yod, Hashim:2020sez, Hashim:2021pkq, Akarsu:2024nas}). The issue of local Lorentz invariance in $f(T)$ gravity has been addressed in \cite{Li:2010cg, Sotiriou:2010mv}. One of the goals of this paper is to better understand the effect of a Lorentz transformation on the properties of $f(T)$ gravity.
Specifically, our study will focus on trying to understand the influence of a background LT on both the background dynamics and the evolution of the perturbations. We will consider two different background spacetimes: a homogeneous and isotropic flat Friedmann-Lemaître-Robertson-Walker (FLRW) and a homogeneous but anisotropic Bianchi I solution endowed with an isotropic two-dimensional spacelike plane. We will find that the large symmetries of FLRW impose the background components of the LT to be time and space-independent, i.e.\ a constant Lorentz transformation. This form of background LT will not affect either the background or the perturbations.
On the other hand, if we consider a homogeneous but anisotropic background, the six components of the background LT will be in general time-dependent; specifically their dynamics can be fixed once their initial conditions and the dynamics of at most two free functions are given, compatibly with the given background symmetries. This implies we have a two-function family of background possibilities. Despite this, we will show that the background dynamics of the $f(T)$ theory are not affected. Moreover, the propagating perturbation variables will not be affected by these redundant free functions for the background solutions either.

One further goal of this paper is to investigate the propagating degrees of freedom (dof) of the $f(T)$ theory of gravity using a perturbative approach, both in a Friedmann-Lemaître-Robertson-Walker (FLRW) spacetime and in a Bianchi I spacetime.
In analogy with $f(R)$ gravity \cite{DeFelice:2010aj,Teyssandier:1983zz}, one would expect that their number increases when considering an action non-linear in the torsion scalar $T$. In the literature several attempts have been made to compute the degrees of freedom of $f(T)$ \cite{Diaz:2014yua, Diaz:2017tmy, Ferraro:2020tqk, GonzalezQuaglia:2023zoc}; some of these works tried to answer this fundamental question using the Hamiltonian formulation of $f(T)$ gravity \cite{Li:2011rn, Ferraro:2018axk, Guzman:2019ozl, Blagojevic:2020dyq}. Interestingly, no agreement has been reached; for instance, the authors of \cite{Li:2011rn} found three propagating dof, the authors in \cite{Ferraro:2018axk} found five, while in \cite{GonzalezQuaglia:2023zoc} they found only two. Another way to compute the number of propagating degrees of freedom consists of using a perturbative approach, i.e.~expanding the vierbein on a specific spacetime. This method is the one we apply in this work; it is not as general as the Hamiltonian approach, however, it sheds some light on the debate. The analysis in FLRW has already been carried out in the literature \cite{Dent:2010nbw, Chen:2010va, Izumi:2012qj, Li:2011wu, Golovnev:2018wbh, Sahlu:2019bug, Bahamonde:2022ohm, Hohmann:2020vcv}. Although the scalar $T$ is a function of the square of first derivatives of the vierbein only,\footnote{In this regard it is more similar to K-essence \cite{Chiba:1997ej,Armendariz-Picon:2000ulo}, for which one has a function $f$ of the kinetic term of a scalar, $f(-\nabla_\mu\phi \nabla^\mu\phi/2)$, rather than $f(R)$, which is instead a function of the second derivatives of the metric tensor.} two degrees of freedom are found to propagate in the gravity sector, corresponding to the two polarizations of the gravitational waves. Hence no new degrees of freedom appear compared to GR. Nevertheless, it could be that the extra degrees of freedom are frozen by the symmetries of the specific spacetime chosen. Therefore in this paper, we perform the same analysis on an anisotropic Bianchi I spacetime. We find that even in this less symmetric spacetime only two degrees of freedom propagate.

The paper is organized as follows. In Section \ref{sec:model} we present a short introduction to $f(T)$ gravity and we develop the formalism we will use in the rest of the article. The role of the Lorentz transformations is highlighted. Section \ref{sec:FLRW} focuses on the study of the background spacetime and the evolution of the perturbations on a FLRW spacetime. The same analysis is carried out for a Bianchi I spacetime in Section \ref{sec:BianchiI}. We conclude and discuss our results in Section \ref{sec:conclusions}. In Appendix \ref{sec:dof_computations} we report in detail the derivation of the propagating degrees of freedom on a Bianchi I spacetime.

\section{The model}
\label{sec:model}

Teleparallel gravity is usually described within the vierbein (or tetrad) formalism, in which the dynamical variable of the theory is not the metric but a four-dimensional vierbein $e^{a}{}_{\mu}$, i.e.\ a collection of four-dimensional covectors, giving an orthonormal basis in the tangent space at each point $x^\mu$ of the manifold.
The Greek letter $\mu$ is a spacetime index, whereas the Latin letter $a$ is an index in the 4-dimensional tangent bundle. In this framework, the metric is expressed in terms of the vierbein itself, as
\begin{equation}
\label{eq:gmn}
    g_{\mu\nu}=\eta_{ab}\, e^{a}{}_{\mu}e^{b}{}_{\nu}\,,
\end{equation}
where $\eta_{a b}$ is the Minkowski metric.

To be as general as possible, we further introduce a general proper orthochronous Lorentz transformation $\Lambda^{A}{}_b$,\footnote{There are two different types of gauge transformations in gravity: local changes of frames and diffeomorphisms. The theory under consideration is diffeomorphism invariant, whereas this work explores the action of a Lorentz transformation, i.e.~a local change of frame.} satisfying
\begin{equation}
\label{eq:LC}
    \eta_{cd} = \Lambda^{A}{}_c\,\eta_{AB}\, \Lambda^{B}{}_d\,.
\end{equation}
Here capital letters refer to Lorentz transformed objects.
Out of these two variables, $e^{b}{}_{\nu}$ and $\Lambda^{A}{}_b$, we can build the most general vierbein as
\begin{equation}
\label{eq:vierbeinL}
    e^{A}{}_{\nu} = \Lambda^{A}{}_b e^{b}{}_{\nu}\,.
\end{equation}
Thanks to the Lorentz condition in Equation (\ref{eq:LC}), one can show that the metric $g_{\mu\nu}$ does not depend on $ \Lambda^{A}{}_b$, i.e.~it is Lorentz-independent:
\begin{equation}
\label{eq:gmnL}
    g_{\mu\nu}=\eta_{ab}\, e^{a}{}_{\mu}e^{b}{}_{\nu}=\eta_{AB}\, e^{A}{}_{\mu}e^{B}{}_{\nu}\,.
\end{equation}
Hence the variables of $\Lambda^{A}{}_b$ do not affect $g_{\mu\nu}$ on any background and at any order in the perturbations. The action for General Relativity, built with the Ricci scalar, is then Lorentz independent too. However, this is not the case for $f(T)$ gravity, where the fundamental object of the theory is the torsion tensor, which is not fully specified by the metric $g_{\mu\nu}$.

We now define the inverse general vierbein as $e^\alpha{}_{A}$, satisfying the properties
\begin{equation}
\label{eq:deltas}
    e^\alpha{}_{A}\, e^{A}{}_\beta = \delta^\alpha{}_\beta\,,\qquad
e^{A}{}_\beta\,e^\beta{}_{B} = \delta^{A}{}_{B}\,.
\end{equation}
Starting from the vierbein, one can also define the curvature-less Weitzenb\"ock connection as:
\begin{equation}
\label{eq:Weizen}
    \Gamma^\alpha{}_{\mu\nu} = e^\alpha{}_{A} \,\partial_\mu e^{A}{}_{\nu}\,,
\end{equation}
Notice that the Lorentz transformed vierbein has been used to build the connection. This is equivalent to setting the connection to be equal to:
\begin{equation}
\label{eq:WeizenL}
    \Gamma^\alpha{}_{\mu\nu} = e^\alpha{}_{a} \,\partial_\mu e^{a}{}_{\nu} + e^\alpha{}_a\, \omega^a{}_{b\mu}\, e^b{}_{\nu}\,,
\end{equation}
where $\omega^a{}_{b\mu}$, called the spin connection, is a pure gauge connection. Indeed, applying a Lorentz transformation to equation (\ref{eq:Weizen}), but defined with the vierbein $e^a{}_\mu$, we get the result in equation (\ref{eq:WeizenL}) with 
\begin{equation}
\label{eq:omega}
    \omega^a{}_{b\mu} = \Lambda^a{}_C\, \partial_\mu \Lambda^C{}_b\,.
\end{equation}
Since it is a pure gauge, the spin connection is usually set to zero when working on a FLRW background. However in this paper, we want to understand the effects of a Lorentz transformation on the properties of $f(T)$ gravity, hence we will keep the general form (\ref{eq:Weizen}). We will show that on a FLRW spacetime, the Lorentz transformation does not affect the results, therefore the spin connection could be safely set to zero already at the beginning. However, this is not the case on a less symmetric spacetime, as the Bianchi I anisotropic spacetime we consider in Section \ref{sec:BianchiI}.

Given the connection, one can define the torsion tensor as its antisymmetric part, i.e.
\begin{equation}
\label{eq:Tamn}
    T^\alpha{}_{\mu\nu}=-2\Gamma^\alpha{}_{[\mu\nu]}\,.
\end{equation}
Notice that the torsion tensor has only spacetime indices.\footnote{It is simple to show that under a general coordinate transformation, $T^\alpha{}_{\mu\nu}$ is a (1,2) tensor, i.e., it transforms according to $T^{\alpha'}{}_{\mu'\nu'}=\frac{\partial x^{\alpha'}}{\partial x^\beta}\,\frac{\partial x^{\sigma}}{\partial x^{\mu'}}\,\frac{\partial x^{\rho}}{\partial x^{\nu'}}\, T^\beta{}_{\sigma\rho}$.}

Starting from the torsion tensor, the torsion scalar can be constructed as:
\begin{equation}
\label{eq:T}
    T = \frac14\,T_{\alpha\mu\nu}T^{\alpha\mu\nu} + \frac12\,T_{\alpha\mu\nu}T^{\mu\alpha\nu}-T^\mu T_\mu\,,
\end{equation}
where the indices have been raised or lowered with the metric $g_{\mu\nu}$ and the contraction $T_\mu$ is defined as $T_\mu = T^\alpha{}_{\alpha\mu}$.
Since the Weitzenb\"ock connection is only a function of the first derivative of the Lorentz transformed vierbein, the scalar $T$ is a quadratic function of the first derivatives. In particular, it does not contain any second derivatives of the fields.

With the torsion scalar, one can construct the action for the gravity sector as:
\begin{equation}
\label{eq:action0}
    S=-\frac{\Mpl^2}{2}\int d^4x \,e\,f(T)\,,
\end{equation}
with $e=\det(e^{A}{}_\mu)=\det(e^{a}{}_\mu) = \sqrt{-g}$.

In the following, we are going to use the Arnowitt–Deser–Misner (ADM) formalism and decompose the metric as:
\begin{equation}
\label{eq:dsADM}
    ds^2=-N^2\,dt^2 + \gamma_{ij}\,(dx^i+N^i dt )\,(dx^j+N^j dt)\,,
\end{equation}
where $N$ is the lapse, $N_i$ is the shift 3D-covector and $\gamma_{ij}$ is the three-dimensional spatial metric. In the following, we will use the Latin indices $i$, $j$ and $k$ for the three-dimensional manifold; they are raised and lowered with the 3D-metric $\gamma_{ij}$. The vierbein has to be written in ADM formalism too.
In general, it can be decomposed as:
\begin{equation}
\label{eq:e_LLe}
   e^{A}{}_{\nu} = \left( \Lambda_{\mathbf{B}} \right)^{A}{}_{\bar b} \left( \Lambda_{\mathbf{R}} \right)^{\bar b}{}_{c} \,e^{c}{}_\mu = \left( \Lambda_{\mathbf{B}} \right)^{A}{}_{\bar b} e^{\bar b}{}_\mu\,,
\end{equation}
where we split the Lorentz transformation $\Lambda^A{}_b$ into $\Lambda_{\mathbf{B}}$, representing the boosts, and $\Lambda_{\mathbf{R}}$, giving the rotations. $e^{\bar b}{}_\mu$ is instead the vierbein after a three-dimensional rotation has been performed. The rotated vierbein $e^{\bar b}{}_\mu$ can be written in ADM formalism as\footnote{See \cite{Faraji:2024enq} for a review of ADM-vierbein formalism.}
\begin{align}
\label{eq:eADM}
    \left\| e^{\bar b}{}_\mu \right\| & = \left(
    \begin{array}{cc}
        N & \vec{0}^{\,T} \\
        e^{\mathcal{I}}{}_i\, N^i  & e^{\mathcal{I}}{}_j
    \end{array}
    \right)\\
    \label{eq:eADM1}
    &=\left(
    \begin{array}{cc}
        1 & \vec{0}^{\,T} \\
        \vec{0}  & O^{\mathcal{I}}{}_\mathscr{j}
    \end{array}
    \right)\left(
    \begin{array}{cc}
        N & \vec{0}^{\,T} \\
        e^{\mathscr{j}}{}_i\, N^i  & e^{\mathscr{j}}{}_j
    \end{array}
    \right)=\left(\Lambda_R\right)^{\bar b}{}_{c} \,e^{c}{}_\mu\,,
\end{align}
where we have introduced the three-dimensional vierbein $e^\mathscr{j}{}_j$ and the rotated three-dimensional vierbein $e^{\mathcal{I}}{}_j$, out of which we can find the three-dimensional spatial metric as
\begin{equation}
\label{eq:gamma_ij}
    \gamma_{ij}=\delta_{\mathcal{IJ}}\,e^{\mathcal{I}}{}_i\,e^{\mathcal{J}}{}_j = \delta_{\mathscr{ij}}e^{\mathscr{i}}{}_ie^\mathscr{j}{}_j\,.
\end{equation}
The indices $\mathcal{I}$ and $\mathcal{J}$, as well as $\mathscr{i}$ and $\mathscr{j}$, are spatial indices on the tangent bundle.
The vierbein $e^\mathcal{I}{}_j$ has been obtained from the three-dimensional vierbein $e^\mathscr{j}{}_j$ applying a proper orthogonal transformation, namely $e^{\mathcal{I}}{}_j=O^{\mathcal{I}}{}_\mathscr{j}\,e^{\mathscr{j}}{}_j$.

Notice that $e^{A}{}_{\nu}$ in \eqref{eq:vierbeinL} has sixteen independent components, while a Lorentz transformation generally depends on six independent spacetime functions (three angles encoded in $\Lambda_{\mathbf{R}}$ and three rapidities encoded in $\Lambda_{\mathbf{B}}$). We can then assume that $e^{c}{}_{\nu}$ has only ten independent components, as the independent elements of a symmetric four-dimensional tensor, e.g.~the metric tensor in \eqref{eq:gmnL}. Four components are given by the lapse $N$ and the shift $N_i$, while the remaining six independent components are encoded in $e^{\mathscr{j}}{}_j$. Without loss of generality, we choose $e^\mathscr{j}{}_j$ to be symmetric. It can be indeed proved that applying a rotation matrix $O^{\mathcal{I}}{}_\mathscr{j}$ on a symmetric matrix one obtains a generic $3 \times 3$ matrix.\footnote{Vice-versa, one can always go from a general $3 \times 3$ matrix, out of an orthogonal matrix, into a symmetric one. Other possibilities exist; for instance, one could have chosen to use a lower (or upper) triangular form instead of a symmetric configuration.}

The decomposition \eqref{eq:e_LLe} and the ADM expression for the vierbein in \eqref{eq:eADM1} will be used in the following calculations.
Having defined the building blocks of the theory, we can further proceed by studying some of its properties.

\subsection{The Role of the Background Values of a Lorentz Transformation}
\label{subsec:backL}

In the previous section, we introduced the fundamentals of $f(T)$ gravity. To fully understand a theory, one needs to study its background solutions, as well as the propagation of the perturbation fields. This analysis will prove the classical stability of the theory in the sub-horizon regime, which is a necessary condition for the quantization of the theory.\footnote{Notice that the absence of a possible infinite strong coupling, another necessary condition for the quantization, is not proved in this work.}
This section is devoted to the analysis of the background solutions of $f(T)$ gravity.

The building block of the theory is the vierbein (\ref{eq:vierbeinL}). One may ask how a Lorentz transformation, and specifically its background value, affects the background solutions (as well as the perturbations dynamics). We have indeed seen in the previous section that, although the metric (\ref{eq:gmn}) is Lorentz independent, the vierbein itself, and consequently the action built with it, is not.
Let us then consider the Weitzenb\"ock connection \eqref{eq:WeizenL}, written in terms of the Lorentz transformed vierbein.
Only the second term is affected by the Lorentz transformation, with the spin connection $\omega^a{}_{b\mu}$ given by \eqref{eq:omega}. If the background value of $\Lambda^{A}{}_c$ is a constant matrix (on some background) then the spin connection vanishes and the Lorentz transformation does not affect the connection and, in turn, the torsion tensor. The dynamics of the background are then not modified.
However, if the background value of $\Lambda^{A}{}_c$ is not constant, it enters the background solutions and might affect the perturbations evolution equations. One may then wonder if there are solutions, in the context of $f(T)$ gravity, that are compatible with the symmetries of the background and at the same time have nontrivial profiles for $\Lambda^{A}{}_c$.

To proceed with the stability analysis for a generic background, let us split both the vierbein and the Lorentz transformation into a background part and a perturbed part:
\begin{equation}
\label{eq:e_split}
    e^b{}_\nu = {\bar e}^b{}_\nu + \delta e^b{}_\nu\,,
\end{equation}
\begin{equation}
\label{eq:Lamba_split}
    \Lambda^{A}{}_b = {\bar \Lambda}^{A}{}_c\,\mathit{\Lambda}^{c}{}_b\,,
\end{equation}
where
\begin{equation}
\label{eq:lambda_pert}
    \mathit{\Lambda}^{c}{}_b = \delta^{c}{}_b + \delta^{c}{}_d\,\eta^{de}\Omega_{eb}
    -\tfrac12\,\delta^{c}{}_d\,\eta^{de}\Omega_{ef}\eta^{fg}\Omega_{gb}+\dots\,.
\end{equation}
The bar over a variable indicates its background value.
The perturbation variables in $\mathit{\Lambda}^{c}{}_b$ are defined via the antisymmetric matrix $\Omega_{ab}$. Notice that to study the linear perturbation theory, we need to expand $\mathit{\Lambda}^{c}{}_b$ up to the second order in the perturbation fields.

To analyze separately the role of the background Lorentz transformation and the perturbations in $f(T)$ gravity, it is useful to rewrite the general vierbein as
\begin{equation}
    e^A{}_\mu = {\bar\Lambda}^A{}_b\mathit{\Lambda}^b{}_c\,e^c{}_\mu = {\bar\Lambda}^A{}_b \mathscr{e}^b{}_\mu\,,
\end{equation}
where $\mathscr{e}^b{}_\mu$ is the vierbein $e^c{}_\mu$ boosted and rotated by an infinitesimal Lorentz transformation. In this way, we isolate the contribution of the background Lorentz transformation. We can then write the Weitzenb\"ock connection \eqref{eq:Weizen} as 
\begin{equation}
\begin{aligned}
\label{eq:Gamma_MLambda}
    \Gamma^\alpha{}_{\mu\nu} &= e^\alpha{}_{A} \partial_\mu e^{A}{}_{\nu}
    =\mathscr{e}^\alpha{}_c {\bar\Lambda}^c{}_A \,\partial_\mu ({\bar\Lambda}^A{}_b \mathscr{e}^b{}_\nu)\\
    &=\mathscr{e}^\alpha{}_b\,\partial_\mu ( \mathscr{e}^b{}_\nu)+
    \mathscr{e}^\alpha{}_c\, {\bar\Lambda}^c{}_A   \,\partial_\mu ({\bar\Lambda}^A{}_b)\,\mathscr{e}^b{}_\nu\,,
\end{aligned}
\end{equation}
where ${\bar\Lambda}^c{}_A$ and $\mathscr{e}^\alpha{}_c$ are the inverse of ${\bar\Lambda}^A{}_c$ and $\mathscr{e}^c{}_\alpha$ respectively. Looking at the second term of \eqref{eq:Gamma_MLambda}, we see that a constant background Lorentz transformation does not affect the Weitzenb\"ock connection and, in turn, $T$. Therefore, in this case, it can be set to the identity without loss of generality. On the contrary, a non-constant Lorentz transformation has to be taken into account carefully, since in principle it could lead to non-trivial consequences.

We now want to construct the background Lorentz matrix $\bar\Lambda^{A}{}_c$. We will consider such a matrix to be an element of the proper Lorentz orthochronous group. Without loss of generality, independently of the background, this matrix can be constructed such that it depends on six independent functions; in particular, it can be written as 
\begin{align}
\label{eq:lambda_back}
    \bar \Lambda =B^{(x)}\bigl(\eta_x\bigr)\,R^{(x)}(\theta_x)B^{(y)}(\eta_y)\,R^{(y)}(\theta_y)B^{(z)}(\eta_z)\,R^{(z)}(\theta_z)\,,
\end{align}
where $B^{(i)}(\eta_i)$ stands for a boost in the $i$-direction with rapidity $\eta_i$, whereas $R^{(i)}(\theta_i)$ corresponds to a rotation about the $i$-axis by an angle $\theta_i$. Notice that all the six parameters $\eta_i = \eta_i(x^\mu)$ and $\theta_i = \theta_i(x^\mu)$ are functions of the coordinates of the background.

The background Lorentz matrix $\bar\Lambda^{A}{}_b$, together with the background values of the vierbein, is used to construct the background torsion tensor as 
\begin{equation}
\label{eq:T_background}
    {\bar T}^\alpha{}_{\mu\nu}=-\bar e^\alpha{}_b \,\partial_\mu ( \bar e^b{}_\nu)+\bar e^\alpha{}_b \,\partial_\nu ( \bar e^b{}_\mu)
    -\bar e^\alpha{}_b \,\bar e^c{}_\nu\bar\Lambda^b{}_{A}
    \,\partial_\mu (\bar\Lambda^{A}{}_c)
    +\bar e^\alpha{}_b \,\bar e^c{}_\mu\bar\Lambda^b{}_{A}
    \,\partial_\nu (\bar\Lambda^{A}{}_c)\,.
\end{equation}
Notice that this tensor has only spacetime indices so it has to share the same symmetry properties as the spacetime background. This characteristic will help us in restricting the possible background solutions only to those that comply with the background symmetries.

We now restrict to homogeneous backgrounds. In this case, it is possible to pick up a coordinate frame on which the observables are time-dependent only. In particular, requiring that the background torsion tensor $\bar T^\alpha{}_{\mu\nu}$ is a function of time only, $\bar \Lambda^{A}{}_c$ depends necessarily on time only too. This in turn implies that $\eta_i(x^\mu) = \eta_i(t)$ and $\theta_i(x^\mu) = \theta_i(t)$. Since the components of $\bar \Lambda^{A}{}_c$ are scalars according to four-dimensional space-time coordinate transformations, they are allowed to have a non-trivial profile, which has to be taken into account when studying the background dynamics. The number of the independent variables $\eta_i$ and $\theta_i$ can be actually reduced by taking into account the symmetries of spacetime (for instance isotropy in the spatial sections). In the following sections, we will specify the spacetime and study the background solutions consequently, as well as the perturbations dynamics.

\section{FLRW spacetime}
\label{sec:FLRW}

\subsection{The Role of the Background Lorentz Transformation}
\label{sec:LorentzFLRW}

We start our analysis by considering a FLRW spacetime. We first focus on the background solutions; the evolution of the perturbations on this background will be studied in the following section.
We work with the vierbein \eqref{eq:vierbeinL}, where $e^a{}_\mu$ is written in ADM formalism as in the second matrix of \eqref{eq:eADM1}. Each contribution in \eqref{eq:vierbeinL}, i.e.~the Lorentz matrix and the vierbein $e^a{}_\mu$, is split into background and perturbations, as in \eqref{eq:e_split} and \eqref{eq:Lamba_split}. At the background level, we consider the most general homogeneous and isotropic background with zero spatial curvature in four dimensions. Hence we set:\footnote{Note that thanks to time reparametrization invariance, we could also set from the beginning $N(t)=1$ or $N(t)=a(t)$.}
\begin{equation}
    \bar\Lambda^{A}{}_a=\bar\Lambda^{A}{}_a(t)\,,\qquad
    \bar N=N(t)\,,\qquad
    \bar N_i=0\,,\qquad \bar N^i= a^{-2}(t) \delta^{ij} N_j=0\,,\qquad
    \bar e^{\mathscr{i}}{}_j=a(t)\,\delta^{\mathscr{i}}{}_j\,,
\end{equation}
with the bar denoting the background variables.
The background Lorentz matrix is decomposed as in \eqref{eq:lambda_back}. The number of independent components can now be reduced exploiting the symmetries of spacetime. In particular, one should impose at the level of the background $\bar T^0{}_{0i}=0$. Indeed, since the spacetime is isotropic, the background torsion tensor has to be invariant under three-dimensional rotations; since the elements $\bar T^0{}_{0i}$ are changed by this transformation, one has to set them to zero. These three constraints reduce the number of equations of motion (eom) for $\eta_i(t)$ and $\theta_i(t)$; in particular, one finds that the three $\dot{\eta}_i$ are not independent and can be expressed in terms of $\eta_i$, $\theta_i$, and $\dot\theta_i$.\footnote{Here a dot denotes a derivative with respect to the cosmic time $t$.} Imposing these conditions, we find that the relations $T_i=T^\mu{}_{\mu i}=0$, necessary for the isotropy of spacetime, are automatically satisfied.

Let us now consider the elements $T^i{}_{0j}$. Under a constant general three-dimensional rotation, they transform according to $T^{i'}{}_{0'j'}=R^{i'}{}_i \, R^j{}_{j'} T^{i}{}_{0j}$, or in matrix notation $\mathbf{T}^\prime = \mathbf{R}^{-1}\mathbf{T} \mathbf{R}$. On an isotropic background, necessarily $\mathbf{T}^\prime = \mathbf{T}$. This requirement is fulfilled only if $\mathbf{T}\propto \mathbf{I}$. Imposing the off-diagonal terms of $T^i{}_{0j}$ to vanish leads to three new conditions, which further reduce the number of independent equations of motion. In particular, one can check that $\dot \theta_i =0$ $\forall i$. Substituting this result into the expressions for $\dot \eta_i$, it is easy to prove that $\dot \eta_i =0$ $\forall i$, independently of the values of $\eta_i$ and $\theta_i$.
As already underlined in Section \ref{subsec:backL}, a constant Lorentz transformation does not affect the evolution of both the background and the perturbation variables; we then set for simplicity $\bar{\mathbf{\Lambda}}=\mathbf{I}$.\footnote{Notice that other terms, such as $\bar T^{0}{}_{ij}=0$, vanish by default on the background and do not lead to any further constraint.}

\subsection{Background Dynamics and Evolution of the Perturbation Variables}
\label{sec:pertFLRW}

Having fixed the background LT, we now want to study the evolution of the perturbation variables. To make a comparison with the literature, we add to the gravity action \eqref{eq:action0} the action $S_m$ for the matter fields and the cosmological constant $\Lambda$. The action we are going to study is then
\begin{equation}
\label{eq:action}
    S=-\frac{\Mpl^2}{2}\int d^4x \,e\,\left[ f(T) +2\Lambda \right]+ S_m\,.
\end{equation}
The vierbein in \eqref{eq:e_split} is written in ADM formalism; the lapse $N$ and the shift $N^i$ are expanded as
\begin{equation}
    N=N(t)\,(1+\phi)\,,\qquad N^i=\frac{N}{a^2}\,\delta^{ij}\,\partial_j n_s+\frac{N}a\,\delta^{ij}\,n^V_j\,,
\end{equation}
while the three-dimensional vierbein $e^\mathscr{j}{}_j$ in \eqref{eq:eADM1} is decomposed into two scalar, two vector, and one tensor components, as:\footnote{The tensor components satisfy the traceless and transverse conditions.}
\begin{align}
    e^\mathscr{j}{}_j = 
    \delta^{\mathscr{i}k}\left[a\,\delta_{kj}\,(1+\psi) + \partial_{k}\partial_j E/a
    +\partial_{(k}^{\phantom{V}} C^V_{j)}
    +\cfrac{1}{2\sqrt{2}}\,a\, h_{kj} \right],
\end{align}
Here ``V" represents an intrinsic vector perturbation satisfying the transversality condition, $\delta^{ij}\partial_i C^V_j=0$. The vierbein components $e^{a}{}_\mu$ at linear order in the perturbations are then given by:
\begin{align}
    e^{0}{}_0 &= N(t)\,(1+\phi)\,,\\
    e^{0}{}_i &= 0\,,\\
    e^{\mathscr{j}}{}_j &= 
    a (1+\psi )\,\delta^\mathscr{j}{}_j+ \delta^{\mathscr{j}k}\,[\partial_k\partial_j E/a +\partial_{(k}^{\phantom{V}} C^V_{j)}
    +h_{kj}]\,,\\
    e^{\mathscr{j}}{}_0 &=e^{\mathscr{j}}{}_j N^j= \delta^{\mathscr{j}k}\,[N/a\, \partial_k n_s+N n^V_k] + \mathcal{O}(\delta^2)\,,
\end{align}
where $\mathcal{O}(\delta^2)$ stands for quadratic terms in the perturbations. The component $e^{\mathscr{j}}{}_0$ will be necessarily contracted with another vector in the action. On a FLRW background, the quadratic terms in $e^{\mathscr{j}}{}_0$ are thus irrelevant, since there are no vectors propagating at the level of the background. However, when considering the Bianchi I spacetime, such terms have to be taken into account. By construction, the vierbein $e^c{}_\mu$ has ten independent perturbation variables, as already mentioned at the end of Section \ref{sec:model}.

Concerning the infinitesimal Lorentz transformation \eqref{eq:lambda_pert}, we expand the real antisymmetric matrix $\Omega_{ab}$ as
\begin{align}
    \Omega_{0i} &= \partial_i\zeta_s+\zeta^V_i\,, \label{eq:infL0i}\\
    \Omega_{ij} &= \partial_i \tilde{s}^V_j-\partial_j \tilde{s}^V_i+\epsilon_{ijk}\partial_k \bar s\,. \label{eq:infLij}
\end{align}
Notice that a term like $\delta_{ij}\nabla^2$ has not been considered, since the matrix $\Omega_{ij}$ is antisymmetric. The total vierbein is then obtained by the multiplication $e^A{}_\mu = \mathit{\Lambda}^{A}{}_b\, e^b{}_\mu$.

Having specified the perturbation variables for the gravity sector, we now have to study the matter sector. We model the matter by a perfect fluid endowed with the following action
\begin{equation}
    S_m = -\int d^4x\,e\,[\rho(n) + J^\mu \partial_\mu\varphi]\,,
\end{equation}
where $n\equiv\sqrt{-J^\mu J^\nu g_{\mu\nu}}$ is the particle number density, and $J^\mu$ is the vector density giving the fluid four-velocity as $u^\mu=J^\mu/n$. The quantities here introduced are expanded into background and perturbation variables as
\begin{align}
    J^\mu \partial_\mu&=\frac{J^0(t)}{N(t)}\,(1+\delta J^0)\,\partial_t + \frac{1}{a^2}(\delta^{ij}\partial_j \delta j)\,\partial_i + \frac1a\,\delta J_V^i\,\partial_i\,,\\
    \varphi &= \varphi(t)+\delta\varphi\,.
\end{align}

\subsubsection{The Background Equations of Motion}
\label{sec:backFLRW}
Although the equations for the background variables have been already derived in the literature, we report them here for completeness. On a flat FLRW spacetime, the background torsion scalar reduces to
\begin{equation}
\label{eq:TbarFLRW}
    \bar{T}(t)= 6H^2\,,
\end{equation}
providing a useful relation between the background torsion scalar and the Hubble factor $H(t)=\dot{a}/(aN)$. From the background equations of motion, one can easily deduce the modified Friedmann equations
\begin{align}
    \label{eq:Friedmann1FLRW}
    3H^2 &=\frac{\rho}{\Mpl^2} + \Lambda +\frac{f}{2}-T f_{,T}+\frac{T}{2}\,.\\
    \label{eq:Friedmann2FLRW}
    -\frac{2 \dot{H}}{N}-3 H^{2}&=  \frac{p}{\Mpl^2}-\Lambda  +\left(4 T f_{,TT}+2 f_{,T}-2\right)\frac{ \dot{H}}{N}+ T f_{,T}-\frac{f}{2}-\frac{T}{2}\,,
\end{align}
with $f_{,T} = df(T)/dT$ and $f_{,TT} = d^2f(T)/dT^2$. The previous equations have been written in this form to emphasize the modifications to the GR equations. Notice that in the limit $f(T)=T$ one recovers the standard Friedmann equations, as one would expect. Indeed, the equivalence between TEGR and GR has been already proved in the literature \cite{BeltranJimenez:2019esp}. The terms after the cosmological constant on the r.h.s.\ of \eqref{eq:Friedmann1FLRW} and \eqref{eq:Friedmann2FLRW} could be recast into an effective energy density and an effective pressure respectively:
\begin{align}
    \frac{\rho_T}{\Mpl^2} &=+\frac{f}{2}-T f_{,T}+\frac{T}{2}\,.\label{eq:rhoTFLRW}\\
    \frac{p_T}{\Mpl^2} &=+\left(4 T f_{,TT}+2 f_{,T}-2\right)\frac{ \dot{H}}{N}+ T f_{,T}-\frac{f}{2}-\frac{T}{2}\,.\label{eq:pTFLRW}
\end{align}
Both $\rho_T$ and $p_T$ vanish when $f(T)=T$. The effective energy density and pressure satisfy the same continuity equation as for the standard content of the Universe, i.e.
\begin{equation}
\label{eq:continuityFLRW}
    \frac{{\dot\rho}_T}{N}+3H(\rho_T+p_T)=0\,.
\end{equation}

As for the equations of motion for the matter fields, they provide the standard relations
\begin{align}
    J^0(t)&=\bar{n}=\frac{{\mathcal{N}_0}}{a^3}\,, \label{eq:J0}\\
    \varphi(t)&=-\int^t dt' N(t'){\bar\rho}_{,{\bar n}}\,, \label{eq:varphi}
\end{align}
where $\bar n$ is the background number density and $\mathcal{N}_0$ is the particle number, having set the value of the scale factor today to unity, i.e.\ $a_0=1$.

\subsubsection{Propagating Degrees of Freedom}
\label{sec:dofFLRW}

Having the action in terms of background and perturbation variables, we can expand it up to second order in the perturbations. To simplify the calculations we choose a specific gauge, the spatially flat gauge.\footnote{This gauge is allowed only if $H\neq0$, as we are going to assume here.} It corresponds to selecting spatial hypersurfaces on which the induced three-dimensional metric is left unperturbed by scalar and vector perturbations. At the level of the perturbations, it amounts to setting 
\begin{equation}
\label{eq:FLRW_gauge}
    C^V_i =0\,,\qquad \psi = 0 = E\,.
\end{equation}

With this gauge choice, the torsion scalar $T$ can be easily computed and inserted into the action \eqref{eq:action}. It turns out all the perturbation variables introduced at the beginning of Section \ref{sec:pertFLRW} enter in the action, except for the field $\bar s$, defined in the infinitesimal Lorentz transformation \eqref{eq:infLij}. This variable disappears completely from the action. Once the action at second order is computed, we move to Fourier space. The convention used for the Fourier transform is the following:\footnote{Notice that, thanks to the symmetry of the system, one is free to choose the perturbations to propagate along one specific direction without loss of generality. We set the x-axis as the direction of propagation to simplify the computations.}
\begin{equation}
\label{eq:fourier_convention}
    \phi(t,x)=\frac1{2(2\pi)^{3/2}}\int_{-\infty}^{\infty} d^3k\, \tilde \phi(t,k)\, e^{i \vec{k}\cdot\vec{x}} + h.c.\,.
\end{equation}
From now on, we will omit both the tilde and the $k$-dependence from the fields for brevity of notation.

Starting from the action at second order in Fourier space, the equations of motion for all the perturbation fields have been derived. Many of these equations are algebraic in one or more variables so that they can be solved easily. The fields specified are thus just Lagrange multipliers, and do not have dynamics. To analyze the dynamical variables, we study separately the tensor, vector, and scalar components, since their equations of motion are indeed decoupled.

The equations of motion for the vector modes are coupled, though all algebraic. It can be shown that their solution leads to vanishing vector modes. For instance, the field $\delta J_V^i$ can be expressed in terms of $n_j^V$ as\footnote{Notice that this expression does not depend on the $f(T)$ theory, as it comes purely from the matter Lagrangian.}
\begin{equation}
    \delta J_V^i = -\bar{n} \,\delta^{ij} n_j^V\,.
\end{equation}
However, solving the equations of motion for $n_j^V$ gives $n_j^V=0$ identically, so that $\delta J_V^i=0$ too. A similar reasoning applies for $\zeta_i^{V}$ and $\tilde s_i^V$. In particular, the equation of motion for $\tilde s_i^V$ sets the constraint
\begin{equation}
    f_{,TT}\,\zeta^V_i=0\,,
\end{equation}
which leads to $\zeta^V_i=0$. Substituting this result back into the eom for $\tilde s_i^V$ leads directly to $\tilde s_i^V=0$. Notice that for $f(T)=T$, the field $\zeta^V_i$ is not constrained but disappears automatically from the action as it happens for a gauge mode.

Concerning the scalar modes, the first equation of motion that has been solved is the one for $\delta j$, which can be expressed in terms of the perturbation $\delta \varphi$, the background number density $\bar n$, given in \eqref{eq:J0}, and the background energy density $\bar \rho$:
\begin{equation}
    \delta j = -\bar{n} n_s +\frac{\bar{n}\,\delta\varphi}{\bar{\rho}_{,{\bar n}}}\,.
\end{equation}
Solving then the equation of motion for $\delta \varphi$ one finds
\begin{equation}
    \delta\varphi= \frac{a^2}{k^2}\,\bar{\rho}_{,\bar{n}}\,\frac{\dot{\delta J}^0+\dot{\phi}}{N}+\bar{\rho}_{,\bar{n}}\,n_s\,.
\end{equation}
The equation of motion for $\zeta_s$ leads instead to:
\begin{equation}
    \zeta_s = \frac{n_s}{a}-\frac{3Ha\phi}{k^2}\,.
\end{equation}
At this point, it is useful to perform a field redefinition. Let us start by introducing
\begin{equation}
    \delta \equiv \frac{\delta\rho}{\bar \rho}\,,
\end{equation}
with $\delta\rho=\rho-\bar\rho$. The right-hand side of the previous equation can be expanded up to linear order in the perturbations, leading to the following field redefinition
\begin{equation}
    \delta J^0 = \frac{\bar\rho}{{\bar n}{\bar \rho}_{,{\bar n}}}\,\delta -\phi\,.
\end{equation}
The equation for $n_s$ can now be solved for $n_s$ in terms of the new field $\delta$ and the remaining scalar perturbation $\phi$:
\begin{equation}
    n_{s} = -\frac{a^{2} \rho  \dot{\delta}}{\bar{n}\bar{\rho}_{,\bar{n}} k^{2} N }-\frac{3 H a^2 \left(\bar{n}\bar{\rho}_{,\bar{n}\bar{n}} \bar{\rho}  -\bar{n}\bar{\rho}_{,\bar{n}}^{2} +\bar{\rho}  \bar{\rho}_{,\bar{n}} \right) \delta}{k^{2} \bar{n}\bar{\rho}_{,\bar{n}}^{2}}-\frac{2\Mpl^2  H f_{,T} \phi}{\bar{n}\bar{\rho}_{,\bar{n}}}\,.
\end{equation}
The eom for the field $\phi$ shows that $\phi$ is a Lagrange multiplier too, and its expression is fully determined by the density perturbation $\delta$. The remaining eom for $\delta J^0$ gives instead the equation for the dynamical field $\delta$. The no-ghost condition for the field $\delta$ leads to the constraint 
\begin{equation}
    \bar{\rho}_{,\bar{n}}>0\,,
\end{equation}
while the speed of propagation can be deduced from its eom to be
\begin{equation}
    c_s^2=\frac{\bar{n}\bar{\rho}_{,\bar{n}\bar{n}}}{\bar{\rho}_{,\bar{n}}}>0\,.
\end{equation}
These results are the same as those obtained in General Relativity. However, the growth of the perturbations is affected by the modification of gravity. Let us consider for instance the case of a dust fluid. In this case $\rho\propto n$, hence $c_s^2=0$. In the sub-horizon limit the eom for $\delta$ reduces to
\begin{equation}
    \ddot \delta + 2 H \dot \delta - \frac{\rho}{2 \Mpl f_{,T}}\delta =0\,.
\end{equation}
It is then immediate to check that the dynamics of the perturbations are now governed by the effective gravitation potential $G_{\rm eff}$, with
\begin{equation}
    \frac{G_{\rm eff}}{G_{\rm N}} = \frac{1}{f_{,T}}>0\,.\label{eq:bary}
\end{equation}

Regarding the two tensor modes, they are decoupled from all the other fields, and are both dynamical, giving the two polarizations of the gravitational waves. The no-ghost condition for the tensor modes gives the constraint
\begin{equation}
    f_{,T}>0\,,
\end{equation}
whereas their speed of propagation in the sub-horizon regime is equal to the speed of light.

In summary, although the dynamics of the propagating degrees of freedom are affected by the modification of gravity, their number on the homogeneous and isotropic FLRW spacetime is three - the density perturbation, and the two polarizations of the GW - as it is in GR.

\section{Bianchi I Spacetime}
\label{sec:BianchiI}

\subsection{The Role of the Background Lorentz Transformation}
\label{sec:LorentzBianchiI}

We now consider an anisotropic Bianchi I spacetime. Let us start studying the effect of the background Lorentz transformation. To this purpose, we consider the background expression for the vierbein in \eqref{eq:e_split} and for the Lorentz matrix in \eqref{eq:Lamba_split}. $\bar \Lambda^A{}_c$ is again given by \eqref{eq:lambda_back}, while the vierbein components in the most general $1+1+2$ background are set to
\begin{equation}
    \bar N=N(t)\,,\qquad
    \bar N_i=(c(t),0,0)\,,\qquad \bar N^i=\gamma^{ij} \bar N_j\,,\qquad
    \bar{e}^{\mathscr{i}}{}_j=\left( \begin{array}{ccc}
        a(t) & 0 & 0 \\
        0     & b(t) & 0\\
        0 & 0 & b(t)
    \end{array}  \right),
\end{equation}
where $\gamma^{il}\gamma_{lj}=\delta^i{}_j$. In this way, the background metric is given by
\begin{equation}
    ds^2 = -\left( N^2-\frac{c^2}{a^2} \right) dt^2+2 c\, dt\,dx
    +a^2dx^2+b^2\,(dy^2+dz^2)\,,
\end{equation}
with $N$, $a$, $b$, and $c$ being functions of time only, since the spacetime is taken to be homogeneous. Notice that the metric is invariant under two-dimensional rotations in the $(y{-}z)$ plane. In general, one can perform a coordinate transformation of the type
\begin{equation}
    t =\tilde t\,,\qquad x = \tilde x + f(\tilde t)\,,
\end{equation}
with $df/d\tilde t=-c/a^2$, to bring the metric element into the form
\begin{equation}
    ds^2 = - N^2 d\tilde{t}\,^2 + a^2 d\tilde x^2 + b^2\,(dy^2+dz^2)\,.
\end{equation}
A time re-parametrization can then be applied to set the line element to
\begin{equation}
\label{eq:metric_BKG}
    ds^2 = - dt^2+a^2dx^2+b^2\,(dy^2+dz^2)\,.
\end{equation}
In the last expression, $\tilde t$ and $\tilde x$ have been renamed $t$ and $x$ for ease of notation. In the following, we will consider the background as defined by the metric \eqref{eq:metric_BKG}. This is in turn equivalent to set
\begin{equation}
    \bar N(t)=1\,,\qquad
    \bar N_i=0\,,\qquad \bar N^i=0\,,\qquad
    \bar e^{\mathscr{j}}{}_j=\left( \begin{array}{ccc}
        a(t) & 0 & 0 \\
        0     & b(t) & 0\\
        0 & 0 & b(t)
    \end{array}  \right).
\end{equation}

Having defined the background variables, we can now reduce the number of independent components thanks to the symmetry of spacetime, i.e.~isotropy in the two-dimensional space of the $(y{-}z)$ plane. This implies that all the background vectors in this plane have to vanish, that is:\footnote{In the following, the indices $l$, $m$ and $n$ will refer to the two-dimensional spacetime identified by the $(y{-}z)$ plane.}
\begin{align}
    {\bar T}^t{}_{tm}=0\,,\qquad
    {\bar T}^t{}_{xm}=0\,,\qquad
    {\bar T}^x{}_{tm}=0\,,\qquad
    {\bar T}^x{}_{xm}=0\,,\qquad
    {\bar T}^m{}_{tx}=0\,,\qquad\bar{T}_m = 0\,.
\end{align}
These constraints can be solved for
\begin{equation}
\label{eq:dotp}
    \dot{p}_\lambda = \dot{\eta}_x F_i + \dot{\theta}_x G_i\,,
\end{equation}
where $p_\lambda\in\{\eta_y,\eta_z,\theta_y,\theta_z\}$, and $F_i$ and $G_i$ are functions of $\eta_i$ and $\theta_i$ only, and not of their first derivatives. Therefore the dynamics of four out of six variables is uniquely determined from the dynamics of the remaining two, once their initial conditions are specified.

Besides the vector constraints, one has also to take into account the tensor constraints.
For instance the element $T^0{}_{mn}$ under a two-dimensional rotation in the isotropic subspace of the $(y{-}z)$ plane transforms as $T^{0'}{}_{m'n'}=R^{m}{}_{m'}R^{n}{}_{n'}T^{0}{}_{mn}$, or in matrix notation $\mathbf{R}^T\mathbf{T}\mathbf{R}$. Since $\mathbf{R}$ must be a symmetry, we require that on the isotropic background $\mathbf{T}=\mathbf{R}^T\mathbf{T}\mathbf{R}$. This condition is respected if $T^{0}{}_{mn}=A_1(t)\,\delta_{mn}+A_2(t)\,\epsilon_{mn}$. Since the torsion tensor is antisymmetric, $A_1(t)=0$ by default. Notice that the antisymmetric term $A_2(t)\,\epsilon_{mn}$ is now allowed because the subspace is only two-dimensional; indeed there is only one independent proper rotation acting on this subspace, contrary to the previous FLRW case where the independent rotations were three.\footnote{Another way to see this is to write the torsion in the basis for the 2-forms in a bi-dimensional space, i.e.~$T^0{}_{mn} \mathbf{d} x^m \wedge \mathbf{d} x^n$. One can then check that the basis is invariant under a rotation on the spatial section $(y{-}z)$, given by $t'=t$, $x'=x$, $y'=y\cos\theta_0 + z\sin\theta_0$ and $z'=-y\sin\theta_0+z\cos\theta_0$, that is $\mathbf{d}y'\wedge \mathbf{d} z'=\mathbf{d}y\wedge\mathbf{d} z$. The invariance also holds for the tensor basis $\frac{\partial}{\partial y'}\otimes {\bf d}z'-\frac{\partial}{\partial z'}\otimes {\bf d}y'$.} An equal splitting holds also for the elements $T^{x}{}_{mn}$.
Concerning the elements $T^{m}{}_{0n}$, they transform as $T^{m'}{}_{0'n'}=R^{m'}{}_{n}R^{n}{}_{n'}T^{m}{}_{0n}$, or in matrix notation $\mathbf{R}^{-1}\mathbf{T}\mathbf{R}$. Since $\mathbf{R}$ is a symmetry, one has to require that $\mathbf{T}=\mathbf{R}^{-1}\mathbf{T}\mathbf{R}$. This is possible again when $T^{m}{}_{0n}=A_3(t)\,\delta^m{}_{n}+A_4(t)\,\delta^{ml}\epsilon_{ln}$. A similar expression has to be imposed also for the components $T^{m}{}_{xn}$. 
However, it turns out that all these tensor constraints do not further restrict the number of dynamic variables. Indeed, the elements of $T^{m}{}_{xn}$ are trivially vanishing, whereas $T^{m}{}_{tn}$ is already in the form of a diagonal matrix plus an antisymmetric one. Also the terms $T^m{}_{nl}$ vanish identically on the background, without adding new constraints.

In summary, after having exploited the isotropy in the two-dimensional plane $(y{-}z)$, we are left with two free functions of time, $\eta_x$ and $\theta_x$, determining the dynamics of all the six dynamical functions $\eta_i$ and $\theta_i$ through equation \eqref{eq:dotp}. In principle, these functions can play a non-trivial role in the evolution of the background and the perturbation variables.

\subsection{Evolution of the Perturbation Variables}
\label{sec:pertBianchiI}

As a second step, we need to discuss the perturbation variables. In the Bianchi I spacetime here considered we work with the gravity sector only for simplicity, i.e.~we consider the action \eqref{eq:action0}. The vierbein is again written in ADM formalism, with the lapse and shift expanded as
\begin{equation}
    N=1+\phi\,,\qquad N^i=\bigl(\partial_x n_1/a^2,(\partial_m n_2+b\,n^V_m)/b^2\bigr)\,.
\end{equation}
while the three-dimensional vierbein is written as:
\begin{align}
    \left\| e^{\mathscr{j}}{}_j\right\| = \left(\begin{array}{cc}
        a\,(1+d_1) & \partial_n\partial_x d_2 + \partial_x w^V_n \\
        \delta^{\mathscr{m}n}(\partial_n\partial_x d_2 + \partial_x w^V_n) &  \delta^{\mathscr{m}m}[\,\delta_{mn}\,b\,(1+\psi) + \partial_{m}\partial_n E
        +\partial_{(m} C^V_{n)}]
    \end{array}\right).
\end{align}
Here every two-dimensional tensor has been decomposed into a scalar (even) and vector (odd) component; moreover we imposed $e^{\mathscr{j}}{}_j$ to be symmetric, as already discussed at the end of Section \ref{sec:model}.

For the infinitesimal Lorentz transformation \eqref{eq:lambda_pert}, we write the elements of the anti-symmetric matrix $\Omega_{ab}$ as:
\begin{align}
    \Omega_{0x} &= \partial_x\zeta_1=-\Omega_{x0}\,,\\
    \Omega_{0m} &= \partial_m\zeta_2+\zeta^V_m=-\Omega_{m0}\,,\\
    \Omega_{xm} &= \partial_x\partial_m s_2 - \partial_x s^V_m=-\Omega_{mx}\,,\\
    \Omega_{mn} &= \partial_m \tilde{s}^V_n-\partial_n \tilde{s}^V_m\,,
\end{align}
Once again, a divergence term of the type $\delta_{mn}\nabla^2$ does not contribute, being $\Omega_{mn}$ anti-symmetric.

The total vierbein is then obtained as $e^A{}_\mu = {\bar\Lambda}^A{}_b\mathit{\Lambda}^b{}_c\,e^c{}_\mu = {\bar\Lambda}^A{}_b \mathscr{e}^b{}_\mu$.\footnote{Here, we remind the reader that $\mathit{\Lambda}^b{}_c$ is a function quadratic in the $\Omega_{ab}$ variables as in Eq.\ \eqref{eq:lambda_pert}.} Notice indeed that the background Lorentz matrix is now non-trivial, hence it has to be taken into account when studying the dynamics of the system. Combining all the previous expressions, up to the background Lorentz transformation, the vierbein components at first order in the perturbation variables are given by:\footnote{Notice that when constructing the ADM four-dimensional vierbein, the term $e^\mathscr{j}{}_0$ has to be kept till $\mathcal{O}(\delta^2)$, contrary to the FLRW case. We report here just the linear order for brevity of notation.}
\begin{align}
    \mathscr{e}^0{}_0 &= 1+\phi\,, \label{eq:e00}\\
    \mathscr{e}^0{}_1 &= -a\,\partial_x\zeta_1\,,\\
    \mathscr{e}^0{}_m &= -b\,(\partial_m\zeta_2+\zeta^V_m)\,,\\
    \mathscr{e}^1{}_0 &= -\partial_x\zeta_1 + \partial_x n_1/a\,,\\
    \mathscr{e}^1{}_1 &= a(1+d_1)\,,\\
    \mathscr{e}^1{}_m &= \partial_m\partial_x (d_2+b s_2)+\partial_x (w^V_m-b s^V_m)\,,\\
    \mathscr{e}^m{}_0 &= \delta^{mn}\,[\partial_n (n_2/b-\zeta_2)+n^V_n-\zeta^V_n]\,,\\
    \mathscr{e}^m{}_1 &= \delta^{mn}\,[\partial_n\partial_x (d_2-a s_2) +\partial_x (w^V_n + a s^V_n)]\,,\\
    \mathscr{e}^m{}_n &= \delta^{ml}[b (1+\psi)\delta_{ln} + \partial_l\partial_n E +\partial_{(l} C^V_{n)} + 2b\, \partial_{[l} {\tilde s}^V_{n]}]\,.\label{eq:emn}
\end{align}
Having set the perturbation variables, one can proceed with the analysis of the dynamics.

\subsubsection{The Background Equations of Motion}
\label{sec:backBianchiI}

On a Bianchi I spacetime, the background torsion tensor reduces to:
\begin{equation}
\label{eq:TbarBianchiI}
    \bar{T}(t) = 4HL + 2L^2\,,
\end{equation}
where we have introduced two Hubble factors $H=\dot a/a$ and $L=\dot b/b$.
This expression simplifies to $6H^2$ in the homogeneous and isotropic limit, as expected. It should be noted that $\bar T$ does not depend on $\eta_i$ or $\theta_i$, although the anisotropy of the spacetime leaves undetermined their dynamics, as shown in Section \ref{sec:LorentzBianchiI}. This implies that the modified background Einstein equations do not fix the dynamics of these functions. Nevertheless, notice that although the background scalar $\bar T$ and the scalars made out of $g_{\mu\nu}$, invariant under Lorentz transformations by construction, do not depend on $\eta_i$ or $\theta_i$, there are scalar quantities that do depend on them. For example,
\begin{align}
    {\bar T}_{\alpha\beta\gamma}{\bar T}^{\alpha\beta\gamma}&=
    -2H^2-4L^2 + g(\eta_i, \theta_i, \dot{\eta}_x, \dot{\theta}_x)\,,\label{eq:T_abc}\\
    {\bar T}_\alpha {\bar T}^\alpha &= -(H+2L)^2 + h(\eta_i, \theta_i, \dot{\eta}_x, \dot{\theta}_x)\,,\label{eq:T_vect_sq_BKG}
\end{align}
where $g(\eta_i, \theta_i, \dot{\eta}_x, \dot{\theta}_x)$ and $h(\eta_i, \theta_i, \dot{\eta}_x, \dot{\theta}_x)$ are two complicated functions of all the Lorentz variables and the time derivatives of $\eta_x$ and $\theta_x$.
Equations~\eqref{eq:T_abc} and~\eqref{eq:T_vect_sq_BKG} clearly show that $\eta_i(t)$ and $\theta_i(t)$ are not coordinate gauge choices, since they affect background scalar quantities. Furthermore, their independent dynamics may affect the perturbations, hence in principle different $(\eta_i,\theta_i)$ label different realizations of the $f(T)$ theory.

The background equations of motion on a Bianchi I spacetime can be computed from the action \eqref{eq:action0}. They can be written in a simple form as:
\begin{align}
    \dot H + H^2 &= L^2\,,\label{eq:Heq}\\
    \dot L + L^2 &= L H\,,\label{eq:Leq}\\
    \Lambda &= -\frac{1}{2}f +T f_{,T}\,.\label{eq:frd1}
\end{align}
We underline that these equations are not all independent; just two of them are, while the third one can be found as a combination of the other two or their first derivatives. This is a consequence of the Bianchi identities, which hold due to diffeomorphism invariance. Equation \eqref{eq:frd1} can be solved to find the possible expressions for $f(T)$. Two solutions are found: one is a generic form for $f(T)$, together with the condition $T=T_0={\rm constant}$, satisfying $\Lambda=-f(T_0)/2+T_0\,f_T(T_0)$; the other is $f(T)+2\Lambda\propto \sqrt{T}$. Notice that this limited number of solutions has been found in the absence of matter; including it allows for far more shapes for $f(T)$.

Let us start the analysis by focusing on the first case. If the torsion scalar is a constant, then from \eqref{eq:TbarBianchiI} it follows that $T=T_0=4HL+2L^2$. Combining this expression with \eqref{eq:Leq} leads to an equation for the Hubble factor $L$:
\begin{align}
    \dot{L}=\frac{T_0}{4}-\frac{3}{2}\,L^2\,,
\end{align}
which has two possible solutions:
\begin{align}
    L_1(t)&=\frac{\sqrt{6T_0}}{6}\,\tanh\!\left[{\frac{\sqrt{6T_0}}{4}\,(t-t_*)}\right], &\textrm{ if }\ T_0>0\,,\label{eq:L1}\\
    L_2(t)&=-\frac{\sqrt{-6T_0}}{6}\,\tan\!\left[{\frac{\sqrt{-6T_0}}{4}\,(t-t_*)}\right], &\textrm{ if }\ T_0<0\,.\label{eq:L2}
\end{align}
Here $t_*$ is a constant of integration.
Inserting these results back into \eqref{eq:TbarBianchiI} one can compute also the Hubble factor $H$:
\begin{align}
    H_1(t)&=\frac{\sqrt{6T_0}}{12}\left\{ 3 \coth\!\left[{\frac{\sqrt{6T_0}}{4}\,(t-t_*)}\right] -\tanh\!\left[{\frac{\sqrt{6T_0}}{4}\,(t-t_*)}\right]\right\},&\textrm{ if }\ T_0>0\,,\label{eq:H1}\\
    H_2(t)&=\frac{\sqrt{-6T_0}}{12}\left\{ 3 \cot\!\left[{\frac{\sqrt{-6T_0}}{4}\,(t-t_*)}\right] +\tan\!\left[{\frac{\sqrt{-6T_0}}{4}\,(t-t_*)}\right]\right\},&\textrm{ if }\ T_0<0\,.\label{eq:H2}
\end{align}
In the case $T_0>0$, for $t\gg t_*$, one can show that
\begin{equation}
    H_1-L_1=\tfrac14\sqrt{6T_0}/\{\sinh[\sqrt{6T_0} (t-t_*)/4]\cosh[\sqrt{6T_0} (t-t_*)/4]\}\to 0\,.
\end{equation}
This means that for $T_0>0$, in the late time limit, the spacetime tends to be isotropic, leaning towards a de Sitter solution with $H_{\rm dS}=\sqrt{T_0/6}$. Notice indeed that working on a FLRW spacetime and setting the matter content to zero, one would get similarly either $T={\rm constant}$ or $f(T)+2\Lambda\propto \sqrt{T}$ from \eqref{eq:Friedmann1FLRW}. In the case $T={\rm constant}$, the solution obtained for the Hubble factor is indeed $H=H_{\rm dS}={\rm constant}$, i.e.\ a pure de Sitter spacetime.
However, although in the limit $t\to t_*$ the Bianchi I spacetime seems to approach a FLRW spacetime, one has to be careful regarding the Lorentz functions $\eta_i$ and $\theta_i$. Indeed, their dynamics are not determined; to get a fully homogeneous and isotropic background the condition $(\dot{\eta}_i,\dot{\theta}_i)\to(0,0)$ has to be met. If this does not happen, both the background dynamics and the perturbations dynamics should not be considered as those on a FLRW spacetime.

Solving for the scale factors, one finds the following solutions for $a$:
\begin{align}
    a_1(t)&=\frac{a_*\sinh\!\left[{\frac{\sqrt{6T_0}}{4}\,(t-t_*)}\right]}{\cosh\!\left[{\frac{\sqrt{6T_0}}{4}\,(t-t_*)}\right]^{1/3}}\,,&\textrm{ if }\ T_0>0\,,\label{eq:a1}\\
    a_2(t)&=\frac{a_*\sin\!\left[{\frac{\sqrt{-6T_0}}{4}\,(t-t_*)}\right]}{\left\{\cos^2\!\left[{\frac{\sqrt{-6T_0}}{4}\,(t-t_*)}\right]\right\}^{1/6}}\,,&\textrm{ if }\ T_0<0\,,\label{eq:a2}
\end{align}
and the following ones for $b$:
\begin{align}
    b_1(t)&= b_*\left\{\cosh\!\left[{\frac{\sqrt{6T_0}}{4}\,(t-t_*)}\right]\right\}^{2/3},&\textrm{ if }\ T_0>0\,,\label{eq:b1}\\
    b_2(t)&= b_*\left\{ \cos^2\!\left[{\frac{\sqrt{-6T_0}}{4}\,(t-t_*)}\right]\right\}^{1/3},&\textrm{ if }\ T_0<0\,,\label{eq:b2}
\end{align}
with $a_*$ and $b_*$ two constants of integration. Looking at equations \eqref{eq:a1}-\eqref{eq:b2}, one can see that for some values of $t$ the scale factors go to zero, giving rise to spacetime singularities. Furthermore, there are values of $t$ for which the scale factors blow up, giving rise to another type of spacetime singularity.
In the case $T_0>0$, the solution for $a_1(t)$ never diverges, while it vanishes for $t\to t_*$. Meanwhile, $b_1(t)$ is always well defined, never vanishing and going to infinity only for infinite cosmic time. Therefore the case $T_0>0$ is always well defined apart from the singularity in the initial time $t_*$.
On the contrary, for $T_0<0$, the scale factor $a_2(t)$ vanishes whenever the argument of the sine goes to zero, i.e.~$\sqrt{-6T_0}(t-t_*)/4 = n\pi$, with $n\in \mathbb{Z}$; moreover $a_2(t)$ diverges for finite values of the cosmic time, that is when the argument of the cosine vanishes, i.e.~$\sqrt{-6T_0}(t-t_*)/4 = \pi/2 + n\pi$. This implies that evolving in time $a_2(t)$ moves from one type of singularity to another. The solution for $b_2(t)$ instead vanishes for $\sqrt{-6T_0}(t-t_*)/4 = \pi/2 + n\pi$, and it never blows up. Then, also $b_2(t)$ moves from one singularity to another when time passes by. In summary, we have two kinds of singularities: one for which $a_2(t)\to 0$ while $b_2(t)\to t_*$; and one for which $a_2(t)\to \infty$ and $b_2(t)\to 0$.

Notice that when $a_2(t)\to \infty$, its derivatives diverge too. In particular, looking at the Hubble rate $H_2$ in \eqref{eq:H2}, one can easily show that it diverges whenever $a_2(t)\to 0$ or $a_2(t)\to \infty$, i.e.~$\sqrt{-6T_0}(t-t_*)/4 = n\pi$ or $\sqrt{-6T_0}(t-t_*)/4 = \pi/2 + n\pi$. The Hubble rate $L_2$ in \eqref{eq:L2} diverges too, in particular whenever $b_2(t)\to 0$, that is $\sqrt{-6T_0}(t-t_*)/4 = \pi/2 + n\pi$. Therefore, whenever $\sqrt{-6T_0}(t-t_*)/4 = n\pi/2$ either $H_2$ or both $H_2$ and $L_2$ blow up.
Furthermore, the scalar quantities such as the ones in equations \eqref{eq:T_abc} and \eqref{eq:T_vect_sq_BKG} blow up too,\footnote{Notice that different choices of $(\dot{\eta}_x,\dot{\theta}_x)$ correspond to different realizations of the $f(T)$ theory. In some cases, it is even possible to remove the singularities in ${\bar T}_{\alpha\beta\gamma}{\bar T}^{\alpha\beta\gamma}$ and ${\bar T}_{\alpha}{\bar T}^{\alpha}$ in \eqref{eq:T_abc} and \eqref{eq:T_vect_sq_BKG} respectively. For instance, one possible realization of the $f(T)$ theory corresponds to setting $\eta_y = \theta_y = \eta_z = \theta_z = 0$, leaving undetermined the dynamics of $\eta_x$ and $\theta_x$ only. In this case, \eqref{eq:T_abc} and \eqref{eq:T_vect_sq_BKG} reduce to ${\bar T}_{\alpha\beta\gamma}{\bar T}^{\alpha\beta\gamma} = 2\dot{\eta}_x^2 - 4 \dot{\theta}_x^2 -2H^2 -4L^2$ and ${\bar T}_{\alpha}{\bar T}^{\alpha} = \dot{\eta}_x^2 - (H+2L)^2$ respectively. Setting $\dot{\eta}_x=H+2L$ removes the singularity in \eqref{eq:T_vect_sq_BKG}, giving ${\bar T}_{\alpha}{\bar T}^{\alpha}=0$; at the same time, the scalar quantity \eqref{eq:T_abc} reduces to ${\bar T}_{\alpha\beta\gamma}{\bar T}^{\alpha\beta\gamma}=4T_0-4\dot{\theta}_x^2$, which is non-singular for finite $\dot{\theta}_x^2$. Notice, however, that this choice corresponds to shifting the singularity to the spacetime scalar present in $\bar{\Lambda}^{b}{}_a$, namely ${\eta}_x$.} showing the presence of a spacetime singularity.\footnote{Notice that here, by spacetime singularity, we mean any linear, quadratic, or higher power scalar built out of the building blocks of the theory blowing up to infinity. For example, another non-trivial scalar quantity built out of the metric that blows up is the Kretschmann scalar,
$${\bar R}_{\alpha\beta\gamma\delta}{\bar R}^{\alpha\beta\gamma\delta}=4\frac{{\ddot a}^2}{{\dot a}^2}+8\frac{{\ddot b}^2}{{\dot b}^2}+8H^2L^2+4L^4=\frac13\, T_0^2\{2+\sec^4[\sqrt{-6T_0}(t-t_*)/4]\},$$
while other curvature invariants remain finite, such as ${\bar R}=2T_0$, and ${\bar R}_{\mu\nu}{\bar R}^{\mu\nu}=T_0^2$. A detailed study of the spacetime with $T_0<0$ being geodesics complete or not is beyond the scope of this paper.}

The second solution admitted by \eqref{eq:frd1} is $f(T)+2\Lambda\propto \sqrt{T}$. In this case equations \eqref{eq:Heq} and \eqref{eq:Leq} can be solved independently from the specific form of $f(T)$. A direct computation provides the same results as in the case $T={\rm constant}$, where now $T_0$ merely becomes an integration constant. This means that the solutions of this special form of $f(T)$ are not isolated. For physical purposes, we will consider from now on only the case $T={\rm constant}$, with $T_0>0$ and $t>t_*$.

\subsubsection{Propagating Degrees of Freedom}
\label{sec:dofBianchiI}

Having solved the equations for the background, we now proceed to analyze the perturbations dynamics. From the action expanded up to the second order in Fourier space, we can compute the equations of motion. To simplify the computations we set again the spatially flat gauge, which now corresponds to:
\begin{equation}
\label{eq:BianchiI_gauge}
    C^V_m =0\,,\qquad d_2=0\,, \qquad \psi = 0 = E\,.
\end{equation}
Notice that in the limit of isotropic spacetime, these conditions coincide with those in \eqref{eq:FLRW_gauge}.

Having set the gauge, one can proceed to solving the equations of motion, in order to understand which fields are Lagrange multipliers and which are propagating dof. We move to Fourier space, where now\footnote{We are free to choose the direction of propagation of the perturbations without loss of generality on a Bianchi I spacetime too. However, in this case, one has to take into account the anisotropy of spacetime, due to which the freedom is left in the $(y{-}z)$-plane only. We then choose the y-axis as the direction of propagation in this subspace.}
\begin{equation}
    \phi(t,x,y)=\frac1{2(2\pi)^{3/2}}\int_{-\infty}^{\infty} dk\, d^2q\, {\tilde \phi}(t,k,\vec{q}\,)\, e^{i (kx+\vec{q}\cdot\vec{y})} + h.c..
\end{equation}
From now on, we will omit the tilde from the Fourier transformed fields and the dependencies on the momenta for simplicity.
Starting from the action in Fourier space, the equations of motion for all the fields are computed. We underline that now the time-dependent functions $\eta_i(t)$ and $\theta_i(t)$ appear in the equations of motion too.
The details of the computations to find the propagating fields using their equations of motion are shown in Appendix \ref{sec:dof_computations}; we report here the main results. 

On a Bianchi I spacetime, in general scalar, vector, and tensor modes in principle could be coupled. However, it turns out that a sub-set of equations is self-consistent; these are the equations for $n_m^V$, $w_m^V$, and $C_m^V$. Thanks to the gauge condition, one of these equations is redundant, and the other two can be easily solved. In particular, one can express $n_m^V$ in terms of $w_m^V$, which instead propagates. Its equations of motion are given by:
\begin{equation}
\label{eq:eomwV}
\begin{split}
    0 &= \ddot{w}_m^V + \frac{K^2 (3H-2L)+Q^2H + sK^2H +bQ^2 (2L-H)} {\left(K^2+Q^2\right) (1+s)}\,\dot{w}_m^V + 
    (K^2+Q^2)w_m^V\\
    &\quad~ +\frac{H^2 \left(K^2-Q^2\right)-6 H K^2 L +3K^2L^2-L^2 Q^2 -2sL \left(H K^2+ L Q^2\right)}{\left(K^2+Q^2\right) (1+s)}\, w_m^V\,,
\end{split}
\end{equation}
where we have introduced for simplicity the variable $s\equiv b/a$ and the squared of the physical momenta, i.e.~$K^2=k^2/a^2$, and $Q^2=|\vec{q}|^2/b^2$.
In the sub-horizon regime, for which $k\gg aH$ and $q\gg bL$, the friction term in \eqref{eq:eomwV} becomes negligible, so that the eom can be approximated to:
\begin{equation}
\label{eq:eomwV_subH}
    \ddot{w}_m^{V} + \left( K^2 + Q^2 \right) w_m^{V} \approx0\,.
\end{equation}
It is clear from \eqref{eq:eomwV_subH} that the frequency $\omega$ of the plane waves describing the modes in the WKB approximation is given by:
\begin{equation}
\label{eq:disprel}
    \omega^2 = K^2+Q^2 = \frac{k^2}{a^2} + \frac{q^2}{b^2}\,.
\end{equation}
Therefore $w_a^V$ propagates with the speed of light in the sub-horizon limit.

The equations of motion for the remaining degrees of freedom are all coupled together. The majority of them are algebraic in the perturbation fields so that they can be easily solved. First of all, it should be noticed that $s_m^V$ and $\tilde s_m^V$ completely disappear from all the equations of motion and are thus undetermined.
The fields $\phi$, $n_1$ and $n_2$, and a linear combination of $\zeta_1$, $\zeta_2$, $\zeta_m^V$ and $s_2$ can be directly expressed in terms of the only propagating mode, i.e.~$d_1$. Its equation of motion is given by:
\begin{equation}
\label{eq:eomdxx}
\begin{split}
    0&=\ddot{d}_1+\frac{5 H^2 Q^2+H L \left(10 K^2+Q^2\right)-4 K^2 L^2}{Q^2 (H+L)+2 K^2 L}\,\dot{d}_1
    +(K^2+Q^2)\,d_1\\
    &\quad~ +\frac{4 H^3 Q^2+2LH^2 \left(4 K^2 -3  Q^2\right)-16 HL^2K^2+2L^3 \left(4 K^2+Q^2\right)}{Q^2 (H+L)+2 K^2 L}\,d_1\,.
\end{split}
\end{equation}
Once again, taking the sub-horizon limit of \eqref{eq:eomdxx} one gets an equation similar to \eqref{eq:eomwV_subH}; using the WKB approximation leads to the dispersion relation as in \eqref{eq:disprel}, implying a luminal propagation.

For both the propagating modes, their equations of motion do not depend on $\eta_i(t)$ and $\theta_i(t)$, which then do not affect the physical degrees of freedom. Instead, they appear as multiplying coefficients in some of the expressions relating the Lagrange multipliers to the propagating degrees of freedom.

We further notice that both $w_m^V$ and $d_1$ are physical only if the no-ghost condition $f_{, T}>0$ is satisfied. This requirement comes from the kinetic terms in the reduced Lagrangian for the propagating modes only, which can be written in Fourier space as:
\begin{equation}
\label{eq:Lag}
\begin{split}
    \frac{\mathcal{L}}{a^3\Mpl^2}&=\frac{f_{,T} K^{2} Q^{2} \left(1+s\right)^{2}}{4 (K^{2}+Q^{2})}\, (\dot{w}_m^V)^2 - \frac{f_{,T} K^{2} Q^{2} (1+s)}{4 \left(K^{2}+Q^{2}\right)^{2}} \left[3 K^2 L^{2}- K^2 H(2 L s -H+6 L) \right.\\
    &\left. +(1+s) (K^{2}+Q^2)^2 -Q^{2} \left(2 s L^{2}+H^{2}+L^{2}\right)\right]\, (w_m^V)^2\\
    &+\frac{f_{,T} Q^{4} L^{2} s^{2}}{{[2 K^{2} L +Q^{2}( H +L)]}^{2}}\, (\dot{d}_1)^2 -\frac{f_{,T} Q^{4} L^{2} s^{2}}{[2 K^{2} L +Q^{2} (H+L)]^{3}} \,\left[ 2 Q^2 H^2\, (2 H-3 L ) +\right.\\
    &\left. +8 K^2 H L (H-2L)+(K^2+Q^2)(2 K^2 L+Q^2(H+L)) +8 K^{2} L^{3}+2 Q^{2} L^{3} \right]\, (d_{1})^{2},
\end{split}
\end{equation}
where $(w_m^V)^2 = w_m^V(t,-k,-\vec{q}\,)w_m^V(t,k,\vec{q}\,)$ and $(\dot{w}_m^V)^2 = \dot{w}_m^V(t,-k,-\vec{q}\,)\dot{w}_m^V(t,k,\vec{q}\,)$, and analog replacements hold for the field $d_1$. Looking at \eqref{eq:Lag}, it is clear that the no-ghost condition is $f_{, T}>0$.

In conclusion, provided the requirement $f_{, T}>0$ is fulfilled, only two modes (an even mode and an odd mode) propagate on a Bianchi I spacetime. These modes give the two polarizations of the gravitational waves, which propagate at the speed of light in $f(T)$.

\section{Conclusions}
\label{sec:conclusions}

The unknown nature of dark matter and dark energy, alongside unsolved tensions in cosmology, could hint at a modification of the laws of gravity with respect to General Relativity. A detailed study of alternative theories of gravity is then worthwhile to address these inconsistencies of the standard cosmological model. This paper focused on $f(T)$ gravity. In particular, we have highlighted the relevance of the Lorentz matrix transformation on both the background spacetime and the perturbations dynamics; furthermore, we have shed light on the propagating degrees of freedom of this theory on homogeneous, both isotropic and anisotropic, backgrounds.

The vierbein, the fundamental object of $f(T)$ gravity, is defined up to a local (i.e.~spacetime dependent) Lorentz transformation. Although the metric is invariant under this transformation, the torsion tensor constructed with the vierbein in general is not. Understanding how a Lorentz transformation affects the background dynamics and the perturbations is then crucial when studying the theory. In this work, we focused in particular on the influence of a background LT, constructed in \eqref{eq:lambda_back} as the product of three boosts along the three space directions, characterized by rapidities $\eta_i$, and three rotations about the three space directions, characterized by angles $\theta_i$. These six independent functions, which are scalars according to four-dimensional diffeomorphisms, can be constrained by exploiting the symmetries of the background for the tensorial quantity $T^\mu{}_{\alpha\beta}$. Two different background spacetimes have been considered: a flat FLRW spacetime and a Bianchi I spacetime. Both backgrounds are homogeneous, thus restricting the functions to be time-dependent only. We have shown how on the isotropic FLRW spacetime the dynamics of all the $\eta_i$ and $\theta_i$ are frozen. This allows one to set the background Lorentz-transformation to be the identity, without loss of generality. On the contrary, the anisotropy of the Bianchi I spacetime does not fix the dynamics of all the functions; in particular, we have shown that in principle there exists a two-function family of backgrounds characterized by different realizations $(\eta_x, \theta_x)$. The dynamics of the remaining variables, $\eta_{y,z}$ and $\theta_{y,z}$, are indeed uniquely determined from the dynamics of $\eta_x$ and $\theta_x$, once their initial conditions are specified. However, when computing the background value of the torsion tensor \eqref{eq:TbarBianchiI}, and the three modified Friedmann equations \eqref{eq:Heq}, \eqref{eq:Leq} and \eqref{eq:frd1}, it turns out they do not depend on these free functions. Therefore, $\eta_i(t)$ and $\theta_i(t)$ do not affect the background in any possible way.

For the Bianchi I spacetime, the Friedmann equations \eqref{eq:Heq} and \eqref{eq:Leq}, and the cosmological constant, i.e.\ the Hamiltonian constraint, \eqref{eq:frd1} have been solved to find the possible Universe dynamics. In particular, the Hamiltonian constraint \eqref{eq:frd1} allows for two possible solutions: 1) $T=T_0=\rm{constant}$, and 2) $f(T)+2\Lambda\propto \sqrt{T}$. Both the solutions lead to the same expressions for the Hubble rates \eqref{eq:L1}-\eqref{eq:H2} and the scale factors \eqref{eq:a1}-\eqref{eq:b2}. Focusing on the case $T=T_0>0$, we have furthermore shown how in the late time limit the spacetime tends to become isotropic, and in particular a de Sitter spacetime. However, one has to be careful when considering this limit since the dynamics of the Lorentz functions $\eta_i(t)$ and $\theta_i(t)$ are not determined a priori.

In this paper, we have also addressed the issue of the propagating degrees of freedom in $f(T)$ gravity on the studied backgrounds. Although the theory is only built of functions of squares of first derivatives of the fields, i.e.~they satisfy at most second-order partial differential equations, and Ostrogradski ghosts are therefore excluded, it is anyway important to describe the nature and behavior of the degrees of freedom, a topic that is still debated in the literature. The analysis on a flat FLRW spacetime had been already performed in the literature; we have extended the study to a less-symmetric spacetime, to understand if extra degrees of freedom were frozen by the FLRW symmetries of spacetime. The effects of a background Lorentz transformation have also been considered. Our work shows how on a vacuum anisotropic Bianchi I spacetime the propagating degrees of freedom reduce to the two polarizations of the gravitational waves, similar to the flat FLRW scenario. In particular, the speed of propagation of the two polarizations is the speed of light also on a Bianchi I background, as it can be seen in \eqref{eq:omega}; furthermore, the no-ghost condition is still given by $f_{, T}>0$, as it can be proved by studying their reduced Lagrangian \eqref{eq:Lag}. Interestingly, although the LT functions $\eta_i(t)$ and $\theta_i(t)$ can be freely chosen, they do not enter into the equations of motion of the propagating degrees of freedom. They instead link some of the fields that are just Lagrange multipliers to the propagating ones.

Our study suggests intriguing possibilities, paving the way for new interesting routes that require further exploration. The analysis performed in this work could be used to investigate models of anisotropic inflation in the context of $f(T)$ gravity. In this case, the matter content of the theory should be taken into account, and the study of the propagating degrees of freedom should be carefully addressed, since in principle scalar, vector, and tensor modes couple to each other.

Another interesting direction is to look for the propagating modes on inhomogeneous backgrounds. Indeed, our findings suggest that the lack of isotropy does not affect the emergence of new degrees of freedom in $f(T)$ gravity. The absence of extra modes in perturbation theory signals a possible strong coupling problem in the theory \cite{Bueno:2016xff, BeltranJimenez:2020lee, Hu:2023juh}, unless the theory possesses only two degrees of freedom, as advocated by \cite{GonzalezQuaglia:2023zoc}. In \cite{Hu:2023juh} the effective field theory (EFT) approach has been applied on a flat FLRW background to try to address this issue, and an estimate of the strong coupling scale has been provided. However, further studies on different backgrounds are required to confirm this result. Although we did not find extra propagating dof compared to General Relativity, the question remains open. Indeed, it is well known that metric theories of the type $\mathcal{L}=\sqrt{-g} f(R)$ have one additive scalar degree of freedom with respect to GR \cite{Teyssandier:1983zz, Sotiriou:2008rp, DeFelice:2010aj}. One would then expect the non-linear extension of TEGR to have additive dof too. However, the same $f(R)$ theory can be rewritten in terms of the torsion tensor as $\mathcal{L}=\det(e^a{}_\mu) f(T+2\nabla^\mu T_\mu)$ \cite{BeltranJimenez:2019esp}. This Lagrangian introduces higher derivative terms with respect to the standard $f(T)$ Lagrangian \eqref{eq:action0}, implying there could be more degrees of freedom in $f(R)$ compared to $f(T)$. It is then essential to gather more theoretical insight into the propagating dof for $f(T)$ theories of gravity.

\begin{acknowledgments}
    V. D.~acknowledges support from MEYS through the INTER-EXCELLENCE II, INTER-COST grant LUC23115.
    The work of A. D. F. was supported by the JSPS Grants-in-Aid for Scientific Research No.~20K03969.
\end{acknowledgments}

\appendix
\section{Derivation of the Propagating Degrees of Freedom on a Bianchi I Spacetime}
\label{sec:dof_computations}

This appendix is devoted to the derivation of the equations of motion for the two propagating degrees of freedom. Starting from the action \eqref{eq:action0} and using the decompositions \eqref{eq:e00}-\eqref{eq:emn} for the vierbein together with \eqref{eq:lambda_back} for the background Lorentz transformation, one can derive the equations of motion for all the perturbation variables at second order in the action. To simplify the computations, the gauge choice \eqref{eq:BianchiI_gauge} has been done. It turns out that the fields $s_m^V$ and $\tilde s_m^V$ completely disappear from all the equations of motion. Moreover, the eom for $w_m^V$, $n_m^V$, and $C_m^V$ are coupled together, and decoupled from all the other equations. Notice that one of the three equations is redundant since $C_m^V$ is set to zero thanks to the gauge condition. The remaining two equations can be used to fix the dynamics of $n_m^V$ and $w_m^V$. The field $n_m^V$ turns out to be a Lagrange multiplier, that is no time derivatives appear in the Lagrangian. As such, one can use its algebraic equation of motion to integrate it out, as in
\begin{equation}
    n_m^V = \frac{k^2 b \left[a \left(\dot{w_m^V} + (H-2L) w_m^V\right)+b \left(\dot{w_m^V} -L w_m^V\right)\right]}{q^2 a^2+k^2 b^2}\,,
\end{equation}
leaving then only the field $w_m^V$ to propagate in the V-sector, with eom given by \eqref{eq:eomwV}.

The remaining equations are all coupled together. The field $\zeta_1$ is found to be a Lagrange multiplier; in particular, one can solve algebraically its equations of motion to find
\begin{equation}
\label{eq:zeta1}
\begin{split}
    \zeta_1 =& \left\{-[\cosh (2 \eta_z)+\cos (2 \theta_z)] \left[-2 b^2 L \left(a^2 \dot{d_1}- a^2 (2 H+L)\phi+k^2 n_1\right)+q^2 a^2 b (H+L) \zeta_2-q^2 a^2 (H+L) n_2\right]+ \right.\\
    &\left. +2 i q a^2 b \cosh (\eta_y) \cos (\theta_y) \left[\sinh (\eta_z) \sin (\theta_z) \left(k q \dot{\theta_x} s_2+\dot{\eta_x} \zeta_m^V\right)+\cosh (\eta_z) \cos (\theta_z) \left(k q \dot{\eta_x} s_2-\dot{\theta_x} \zeta_m^V\right)\right]+ \right.\\
    &\left. +2 i q a^2 b \sinh (\eta_y) \sin (\theta_y) \left[\sinh (\eta_z) \sin (\theta_z) \left(k q \dot{\eta_x} s_2-\dot{\theta_x} \zeta_m^V\right)-\cosh (\eta_z) \cos (\theta_z) \left(k q \dot{\theta_x} s_2+\dot{\eta_x} \zeta_m^V\right)\right]\right\}\cdot\\
    &\left\{2 k^2 a b^2 L (\cosh (2 \eta_z)+\cos (2 \theta_z))\right\}^{-1}\,.
\end{split}
\end{equation}
Once $\zeta_1$ is fixed with \eqref{eq:zeta1}, the fields $\zeta_2$, $\zeta^V_m$, and $s_2$ completely disappear from the Lagrangian, and their eom vanish consequently. This means that the theory of the linear perturbations only depends on a linear combination of $\zeta_1$, $\zeta_2$, $\zeta_m^V$, and $s_2$. Consequently, the equations of motion for these fields become redundant. At this point, it is easy to prove that also the field $n_1$ is a Lagrange multiplier and can be integrated out using its eom as
\begin{equation}
n_1=n_2-\frac{4 b^{2} L \phi}{q^{2}}\,.
\end{equation}
The equations of motion of $n_2$ can be solved algebraically for $\phi$ giving
\begin{equation}
    \phi=\frac{q^{2} a^{2} b [\left(H -L \right) d_{1}+\dot{d}_{1}]}{a^{2} b \left(H +L \right) q^{2}+2 k^{2} b^{3} L}\,.
\end{equation}
Indeed, the field $n_2$ appears only linearly in the Lagrangian and it multiplies $\phi$; its equation of motion thus sets a constraint on $\phi$ itself. The field $n_2$ can then be completely determined in terms of $\phi$ using the algebraic eom for $\phi$. The only remaining equation is the one for $d_1$, which is exactly given by \eqref{eq:eomdxx}.

\renewcommand\bibname{Bibliography}
\addcontentsline{toc}{chapter}{Bibliography} 
\bibliographystyle{JHEP}
\bibliography{bibliography}

\end{document}